\documentclass{cernrep} 
\usepackage{texnames}
\usepackage[T1]{fontenc}
\usepackage[bookmarks, colorlinks=true, linktoc=page, linkcolor=black, citecolor=black, urlcolor=blue]{hyperref}
\usepackage{cite}

\usepackage{subfigure}
\usepackage{graphicx}
\usepackage{subfigure}
\usepackage{upgreek}
\usepackage{float}

\def\lsim{\mathrel{ \rlap{\raise.5ex\hbox{$<$}}
                      {\lower.5ex\hbox{$\sim$}}  } }
\def\gsim{\mathrel{ \rlap{\raise.5ex\hbox{$>$}}
                      {\lower.5ex\hbox{$\sim$}}  } }

\newcommand{\vect}[1]{\boldsymbol{#1}}
\def\vx{\vect{x}}
\def\vX{\vect{X}}
\def\vy{\vect{y}}
\def\vw{\vect{w}}
\def\vtheta{\vect{\theta}}
\def\xi{x^{(i)}}
\def\yi{y^{(i)}}
\def\vxi{\vect{x}^{(i)}}
\def\ei{\epsilon^{(i)}}
\def\R{\mathbb{R}}
\def\bigO{\mathcal{O}}
\def\p{\vw}
\def\vphi{\vect{\phi}}

\def\tmax {\underset{\theta}{\mathrm{argmax} }\,}
\def\tmin {\underset{\theta}{\mathrm{argmin} }\,}

\usepackage{fancyhdr}
\fancyhfoffset{4 mm}
\fancypagestyle{ARTTITLE}{%
\fancyhf{} 
\lhead{\small{Proceedings of the 2018 CERN--Accelerator--School course on\\
\it{Numerical Methods for Analysis, Design and Modelling of Particle Accelerators}, Thessaloniki, (Greece)}} 
\lfoot{Available online at \url{https://cas.web.cern.ch/previous-schools}}
\rfoot{\thepage\hspace*{3mm}}
 
}

\begin{document}

\title{Introduction to Machine Learning for Accelerator Physics}

\author{D.~Ratner}

\institute{SLAC, Menlo Park, United States}

\begin{abstract}
This pair of CAS lectures gives an introduction for accelerator physics students to the framework and terminology of machine learning (ML). We start by introducing the language of ML through a simple example of linear regression, including a probabilistic perspective to introduce the concepts of maximum likelihood estimation (MLE) and maximum a priori (MAP) estimation. We then apply the concepts to examples of neural networks and logistic regression. Next we introduce non-parametric models and the kernel method and give a brief introduction to two other machine learning paradigms, unsupervised and reinforcement learning. Finally we close with example applications of ML at a free-electron laser.
\end{abstract}

\keywords{Machine learning, AI, neural networks.}

\maketitle
\thispagestyle{ARTTITLE}

\section{Introduction}

This pair of CAS lectures was an introduction for accelerator physics students to the framework and terminology of machine learning (ML).  With the enormous range of ML methods in use, and the rapid pace of change, it is impossible to give a survey of the field in such a brief format.  Instead, the goal of this lecture was to give accelerator students the tools for their own exploration of ML applications to accelerators.

We start by introducing the language of ML through a simple example of linear regression, a familiar subject for most physicists.  We then revisit the regression problem from a probabilistic perspective to introduce the concepts of maximum likelihood estimation (MLE) and maximum a priori (MAP) estimation.  We end this section by applying the concepts to examples of neural networks and logistic regression. Next we introduce non-parametric models and the kernel method.  We end with a brief introduction to two other machine learning paradigms, un-supervised and reinforcement learning.  Finally we close with example applications at a free-electron laser.  The approach we follow here is in part condensed from the~well-known CS229 course at Stanford University \cite{cs229}, available online and highly recommended to the motivated student for more in depth study. 

\section{ML Framework}

\subsection{Linear Regression, machine-learning style} \label{sec:lin_reg}

To introduce the framework of machine learning we start by treating a problem familiar to physicists: linear regression. As with any modeling problem, we start with a data set.  In the language of machine learning, the data is our `training set' consisting of $m$ different examples.  Each of the $m$ examples has a vector of $n$ `features' $\vx$ (the independent variables), and one label $y$ (the dependent variable).  (Note that the labels are also often referred to as the `ground truth.'  We will use these terms interchangeably.) Given a new example, $\vx'$, the goal of our model is to predict the associated label, $y'$. The process of making predictions on new data is sometimes referred to as `inference.'  

Given an example $i$ with features $\vxi$, we will refer to our prediction for the label $\yi$ as the~`hypothesis' $h_\theta(\vxi)$.  In the case of linear regression we have
\begin{align}
h_\theta(\vxi) \equiv \sum_{j=0}^n \theta_j \xi_j  \,,
\label{eq:linreg_sum}
\end{align}
where the $\theta_j$ are the model parameters. Note that the sum is over $n+1$ parameters to allow for an~intercept (or `bias') term, $\theta_0$. By convention we define $x_0 \equiv 1$. Equation (\ref{eq:linreg_sum}) can be written in a more compact form 
\begin{align}
h_\theta(\vxi) = \vxi\cdot\vtheta \,,
\end{align}
with row vector $\vxi \in \R^{1 \times (n+1)}$ and column vector $\vtheta \in \R^{(n+1) \times 1}$.  The learning process can then be stated succinctly as finding the parameter vector $\vtheta$ that produces the best predictions, $h_\theta(\vxi)$.  

\begin{figure}[!htb]
\centering
\includegraphics[width=.6\linewidth]{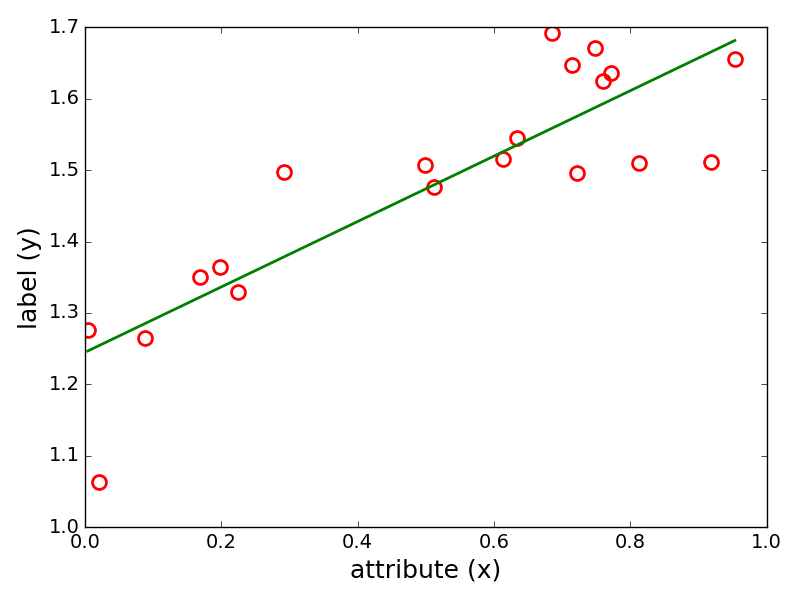}
\caption{A simple linear regression problem.  Data points (red circles) are given for a single feature.  The least squares regression solution is given by the green line.  }\label{fig:linreg}
\end{figure}

For a simple example from accelerator physics, consider calibrating a radiation intensity monitor.  We collect a set of readings from our diagnostic, $\{\vx^{(1)},...,\vx^{(m)}\}$, each corresponding to a known power level, $\{y^{(1)},...,y^{(m)}\}$.  (For example, we might have a second detector that is already calibrated to provide accurate power measurements.) Then given a new reading, $\vx'$, our goal is to predict the corresponding power, $y'$. We pose our task as finding the parameters, $\vtheta$, which minimize the error between our hypothesis $h_\theta(\vx)$ and the known ground truth, $y$. To make the concept of error concrete, we must choose a metric, in ML commonly known as a `cost'  or `loss' function.  The choice of cost/loss function should be given careful consideration, as it can have a strong influence on the resulting model.  As in physics, a common choice is mean squared error (MSE):
\begin{align}
C(\theta) 
&= \frac{1}{2} \sum_{i=1}^m \big(h_\theta(\vxi)-\yi\big)^2 
= \frac{1}{2}(\vX\vtheta-\vy)^T (\vX\vtheta-\vy) \,,
\end{align}
where in compact form $\vX \in \R^{m\times (n+1)}$ and $\vy \in \R^{m}$ are the features and labels for all $m$ examples. The~goal then is to find the values of $\vtheta$ that minimize $C(\vtheta)$, i.e. $\hat{\vtheta} \equiv \tmin C(\vtheta)$. For the special case of the MSE cost function, an analytical solution exists in the form of the normal equations:
\begin{align}
\hat\vtheta = (\vX^T\vX)^{-1} \vX^T\vy \,.
\label{eq:normal}
\end{align}

For general choices of cost functions and models, an analytical solution may not exist.  Alternatively, we can solve for $\hat\vtheta$ by numerical optimization. A common choice is gradient descent: starting from an initial guess, each iteration updates each component $\theta_j$ according to the rule
\begin{align}
\theta_j := \theta_j - \alpha \frac{\partial C(\vtheta)}{\partial \theta_j}  \,.
\end{align}
The parameter $\alpha$ adjusts how aggressively to change $\theta_j$, and thus is known as the learning rate.  In our MSE example we can write down an analytical expression for the partial derivatives,
\begin{align}
\frac{\partial C(\vtheta)}{\partial \theta_j} = \sum_{i=1}^m \big(\yi-h_\theta(\vxi)\big) \xi_j \,.
\label{eq:LS_update}
\end{align}
Equation (\ref{eq:LS_update}) calculates the derivative by averaging over all $m$ examples in the training set for each update of $\vtheta$.  For training sets with many examples, each evaluation may be computationally expensive.  Often it is not necessary to evaluate the entire data set to make a good estimate of the gradient, especially early in the training process.  In the opposite limit, `stochastic gradient descent' updates $\theta_j$ after calculating the derivative for each example.  While more efficient, stochastic gradient descent is sometimes too noisy when gradients are small.  In practice `mini-batch gradient descent,' in which the number of training examples per update is set by the user, an example of a `hyperparameter.' (We will discuss hyperparameters more in the next section.) As training proceeds and the gradient become smaller, increasing the number of examples often leads to best performance.

\subsection{Bias-Variance Tradeoff and Hyperparameters}

We now consider a slightly more complex model.  Suppose we have a single scalar physical input, $x$, and again a scalar label $y$.  This time we will fit a polynomial model
\begin{align}
h_\theta(\xi) \equiv \sum_{k=0}^n \theta_k (\xi)^k \,.
\label{eq:poly_reg}
\end{align}
One way to interpret Eq.~(\ref{eq:poly_reg}) is that we have taken a single physical quantity, $x$, and converted it to $n$ different features by the `feature mapping'
\begin{align}
x \rightarrow \vphi(x) = \{x,x^2,...,x^n\}\,.
\label{eq:feat_def}
\end{align}
The motivation for this terminology will become apparent later in discussion of kernel methods. As physicists, we might call Eq.~(\ref{eq:poly_reg}) polynomial regression, because it is polynomial in the physical quantity, $x$. However, in ML terminology it is still under the umbrella of `linear regression,' because the model is linear in the features, $\vphi(x)$. 

\begin{figure}[!htb]
\centering
\includegraphics[width=.6\linewidth]{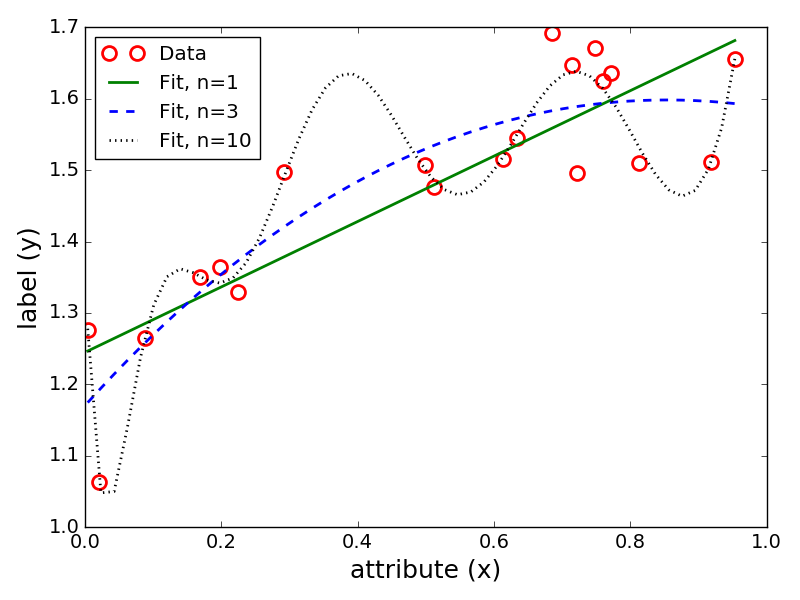}
\caption{Three different choices of $n$ for polynomial regression. The solid green line ($n=1$) has high bias (underfitting), while the black dotted line ($n=10$) exhibits high variance (overfitting). The dashed blue line ($n=3$) appears near an optimal fit.}\label{fig:poly_reg}
\end{figure}

We are now faced with a question: what degree of the polynomial, $n$, in Eq.~(\ref{eq:poly_reg}) is optimal?  The~choice of $n$ is a second example of a `hyperparameter,' i.e. user choices that are not explicit model parameters, $\vtheta$. While we know to estimate $\hat\vtheta$ by minimizing the cost function, how do we select optimal hyperparameters?  Let's work through the case of choosing the polynomial degree. Figure~\ref{fig:poly_reg} shows fits for three different choices, $n=[1,3,10]$. We may intuit that the $n=3$ choice is preferred; the $n=1$ fit appears to miss a physically significant curvature, while $n=10$ appears to be fitting noise rather than the~underlying physics.  We refer to the first case as `high bias' (or under-fitting) because the model is biased to a linear fit, and the second case as `high-variance' (or over-fitting) because the model is capturing variance of the data rather than a true physical relation.

\begin{figure}[!htb]
\centering
\includegraphics[width=.6\linewidth]{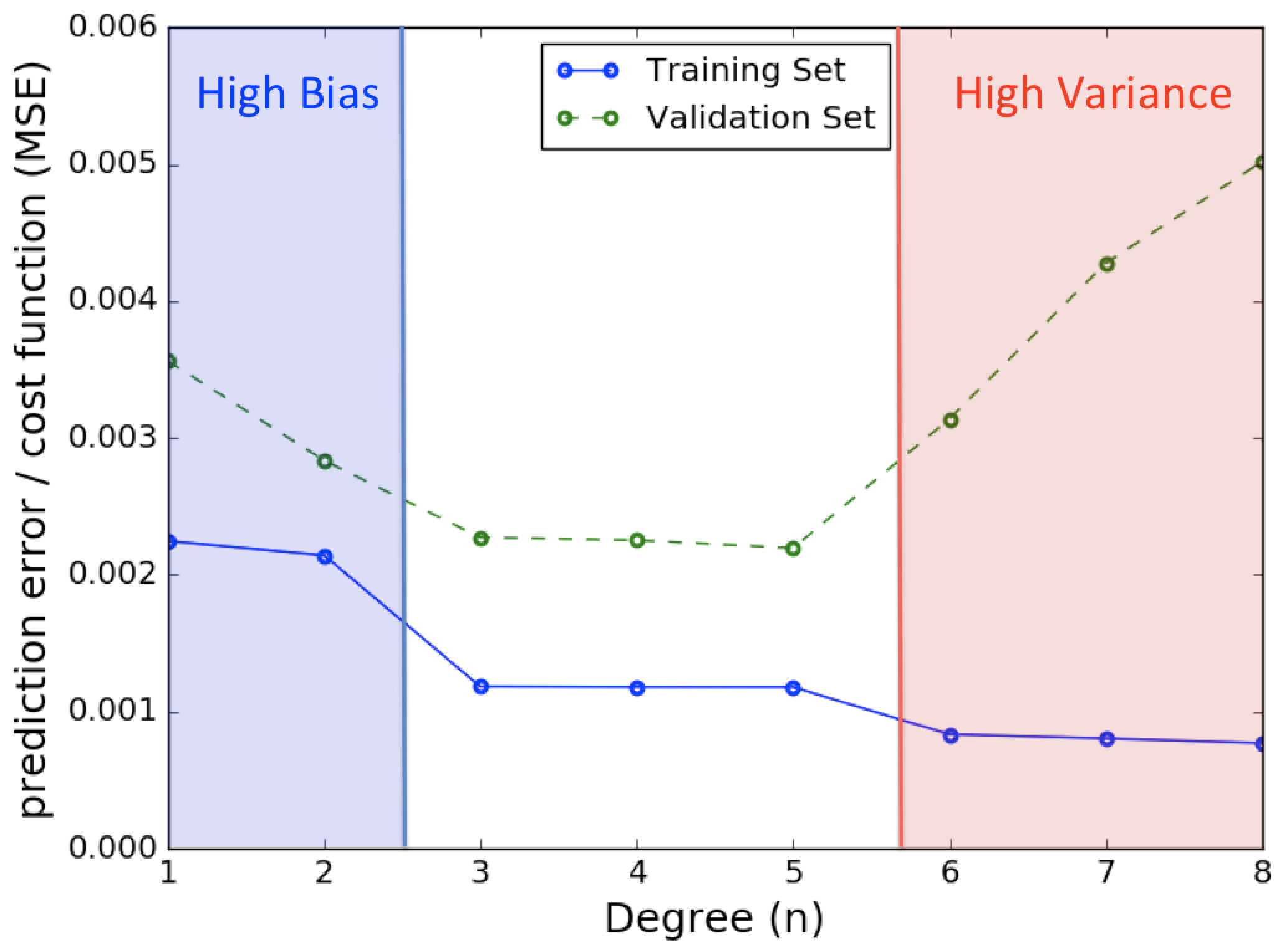}
\caption{Cost values for both the training and validation sets as a function of polynomial degree in Eq.~(\ref{eq:poly_reg}). In the~blue region, the model has high bias and both training and validation costs are high, whereas in the red region the model has high variance, and only the validation cost is large. The best performance is given in the range $n=$3 to $5$, the argmin of the cost function on the validation set.}\label{fig:train_val}
\end{figure}

To make the intuition of the previous paragraph concrete, we introduce the concept of `training' and `validation' data sets. We break the original data set into two components, typically with 80-90\% in the training set and the rest in the validation set. Using the data in the training set, we repeatedly estimate $\hat\vtheta$ for each of the hyperparameter choices.  We then test each model on the examples in the validation set, and select the hyperparameter with the best performance. Figure~\ref{fig:train_val} shows typical behavior.  As the~number of features increases, the training error continues to decrease, but the validation error begins to climb as we start overfitting.  

(Note that whenever reporting performance of a model, it is critical to reserve a third `test' set that is only used a single time at the end of the study.  Repeated optimization of hyper-parameters may lead to overfitting examples in the validation set. The final evaluation score should always use previously unseen data.)

\begin{figure}[!htb]
\centering
\includegraphics[width=.6\linewidth]{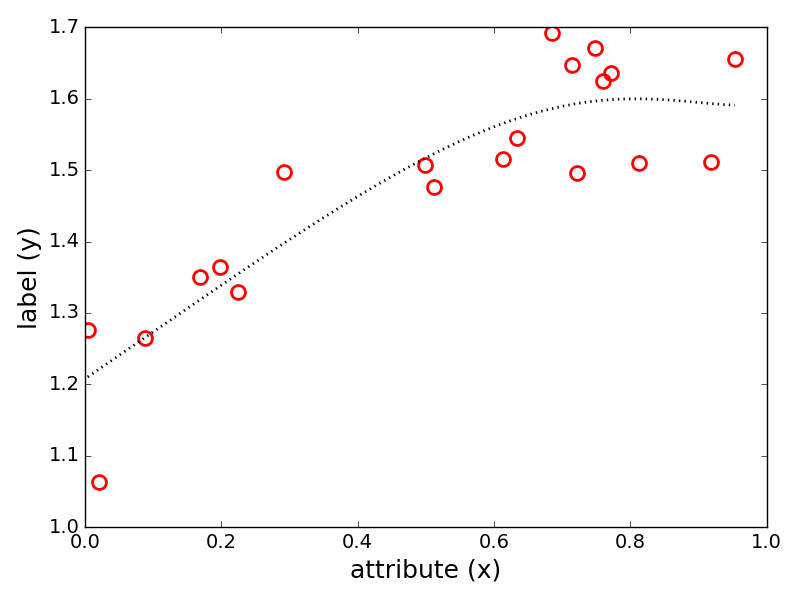}
\caption{Polynomial regression for $n=10$ with L$_2$ regularization.  Compare to the strong overfitting for $n=10$ in Fig.~\ref{fig:poly_reg}.}\label{fig:ls_reg}
\end{figure}

The `bias-variance' trade-off is a central problem for machine learning.  A model that exhibits high bias requires more fitting power, for example through collection of additional types of data or creation of new features. On the opposite side, a high variance model has too much fitting power, and may improve by reducing the number of features (also known as `feature selection').  An alternative solution is the~addition of `regularization' terms to the cost function.  Here we will introduce regularization without formal justification, though we will see it emerge naturally in the next section.  We return to the MSE cost function, now with a new term
\begin{align}
C(\theta) 
&= \frac{1}{2}||h_\theta(\vX)-\vy||_2 + \lambda ||\vtheta||_2 \,,
\label{eq:ridge}
\end{align}
where $||\cdot||_2$ is the L$_2$ norm, and $\lambda$ is the new regularization hyperparameter. (Here the general L$_p$ norm is defined $||\vx||_p \equiv (\sum_i |x_i|^p)^{1/p}$.) Intuitively, increasing the value of $\lambda$ has the effect of encouraging the~individual values of $\theta_j$ to be small; any increase in $\theta_j$ must be offset by an equivalent or larger decrease in the fitting error.  Figure~\ref{fig:ls_reg} shows an example of L$_2$ regularization applied to our linear regression problem.

L$_2$ regularization is appealing because the normal equations (slightly modified) still provide a~closed-form solution.  However, some tasks may benefit from other forms of regularization as well.  For example, the L$_0$ ``norm'' (technically not a proper mathematical norm) is defined as the number of non-zero entries in $\vtheta$; L$_0$ regularization effectively implements feature selection, pushing the model to ignore the least effective features, or equivalently to search for sparse solutions (see e.g. compressed sensing \cite{candes06rt}). While L$_0$ is often computationally impractical (it's NP-hard), the L$_1$ norm produces similar results and is used widely.  As a practical note, in linear regression L$_2$ regularization is often referred to as ridge or Tikhonov regression, L$_1$ regularization is known as LASSO (least absolute shrinkage and selection operator), and the combined L$_1$ and L$_2$ norm is called elastic net.  All are widely available on popular platforms such as Matlab and scikit-learn.

\pagebreak

\subsection{Probabilistic View}

The skeptical reader may question our choice of the MSE metric and L$_2$ regularization simply due to computational convenience. Here we repeat our derivation of linear regression using a probabilistic interpretation; we will see the probabilistic view naturally motivates the choice of metric and regularization.  

We start from the same assumption of a data set with features, $\vX$, and labels, $\vy$. This time we treat both the features and labels as random variables, introducing a random noise term, $\ei$, to give a~new model
\begin{align}
\yi = \vxi\cdot\vtheta  + \ei \,.
\end{align}
If we assume that the noise is normally distributed with zero mean and rms width $\sigma$, then the probability of measuring an outcome $\yi$ given features $\vxi$, and parameterized by $\vtheta$ is
\begin{align}
p(\yi|\vxi;\vtheta) = \frac{1}{\sqrt{2\pi}\sigma} \exp \bigg[ -\frac{(\yi-\vxi\cdot\vtheta)^2}{2\sigma^2} \bigg] \,.
\end{align}
As before, our goal is to pick values of $\vtheta$ that `best' fit this probability distribution.  One logical choice for `best' is to pick $\vtheta$ so that, given a pair of $\vxi,\yi$, we maximize the probability that $h_\theta(\vxi) = \yi$.  This is known as the maximum likelihood estimator (MLE). (Note however that this is not the only possible choice for `best.')  More precisely, we would like to pick $\vtheta$ to maximize the probability over ALL such pairs.  We call the joint probability the `Likelihood' 
\begin{align}
\mathcal{L}(\theta) \equiv \prod_{i=1}^m p(\yi | \vxi; \vtheta) 
= \prod_{i=1}^m \frac{1}{\sqrt{2\pi}\sigma}\exp \bigg[ -\frac{(\yi-\vxi\cdot\vtheta)^2}{2\sigma^2} \bigg] \,.
\label{eq:likelihood}
\end{align}
Dealing with the products is cumbersome. Note that our goal is only to find the argmax of $\mathcal{L}(\vtheta)$, not the~maximum itself, and we are free to apply any monotonic transformation.  In particular, we can take the logarithm of both sides, giving the so-called `log likelihood"
\begin{align}
\ell(\vtheta) \equiv \log \mathcal{L}(\vtheta)
= m \log \frac{1}{\sqrt{2\pi}\sigma} - \frac{1}{2\sigma^2} \sum_{i=1}^m (\yi-\vxi\cdot\vtheta)^2\,.
\label{eq:loglik}
\end{align}
The first term is independent of $\vtheta$ and can be dropped. Applying our MLE principle, $\hat\vtheta \equiv \tmax \, \ell(\vtheta)$, we find that with Gaussian noise
\begin{align}
\hat\vtheta 
= \tmin  \, \frac{1}{2\sigma^2} \sum_{i=1}^m (\yi-\vxi\cdot\vtheta)^2\,.
\label{eq:MLE}
\end{align}
In the end we simply recover least squares regression, or put differently least squares regression is the~result of assuming Gaussian noise and solving with MLE. However, MLE is also a generic approach to fitting model parameters, and can be used for a wide range of assumptions and model types.

Finally, we briefly consider yet a third interpretation, this time using Bayes' rule.  Bayes' rule states that for two random variables, $A$ and $B$,
\begin{align}
p(A|B) &= \frac{p(B|A) p(A) }{p(B)}  \,.
\label{eq:bayes}
\end{align}
(For readers unfamiliar with Bayes, this relation follows directly from the observation of overlap in a~Venn diagram: $p(A | B) = p(A \cap B)/p(B)$ and $p(B | A) = p(A \cap B)/p(A)$.) In Bayesian lingo Eq.~(\ref{eq:bayes}) reads
\begin{align}
\mathrm{ posterior} &= \mathrm{\frac{likelihood \times prior}{evidence}} \,.
\end{align}
The `prior' is our assumed distribution of $A$ before measuring $B$, and the `posterior' is our updated belief after the measurement. Intuitively, Bayes tells us that our prior assumptions can affect our posterior beliefs.  A classic example is a test, $t$, for a rare medical condition, $c$.  Suppose the test only has 1\% false positives and 1\% false negatives, i.e. $p(t=1|c=0)=0.01$ and $p(t=0|c=1)=0.01$.  We also have the prior knowledge that the condition occurs in only 0.1\% of the population: $p(c=1)=0.001$.  What is the probability that a positive result indicates the patient actually has the condition?  Plugging into Bayes formula we find:
\begin{align}
p(c=1|t=1) &= \frac{p(t=1|c=1) p(c=1) }{p(t=1|c=1)p(c=1) + p(t=1|c=0)p(c=0)} \nonumber\\
&= \frac{0.99*0.001 }{0.99*0.001 + 0.01*0.999} \nonumber\\
&\approx 9\%    \,.
\label{eq:bayes}
\end{align}
Despite the seemingly high quality of the test, our prior belief has a strong impact on our posterior confidence in the result. 

Now we apply the Bayesian view to the problem of regression.  The~Bayesian interpretation differs from the previous frequentist view by also treating the model parameters, $\vtheta$, as random variables.  In the~Bayesian view, we restate our goal as finding
\begin{align}
p(\vtheta | \vxi,\yi) = \frac{p(\vxi,\yi|\vtheta) p(\vtheta) }{p(\vxi,\yi)}  \,.
\label{eq:bayes}
\end{align}
Note the denominator (`evidence') has no $\vtheta$ dependence, and for optimization purposes can be ignored.  In practice, solving Eq.~(\ref{eq:bayes}) explicitly is often not computationally feasible, but a common heuristic is maximum a posteriori (MAP) estimation which finds only the most likely value of $\vtheta$ (analogous to MLE)
\begin{align}
\vtheta_{\mathrm{MAP}} = \underset{\vtheta}{\mathrm{argmax} } \prod_{i=1}^m \p(\yi|\vxi,\vtheta) p(\vtheta) \,,
\label{eq:MAP}
\end{align}
where the product is over all examples in the training set. The only difference compared to Eq.~(\ref{eq:likelihood}) is the addition of the~prior term, $p(\vtheta)$; the upshot is that our prior expectation of $\vtheta$, i.e. before training, can affect the final posterior belief after training.  For example, if we believe the values of $\vtheta$ should be small, we set a penalty on using large values of $\vtheta$.  This penalty should sound familiar to the reader; the~prior is a natural way to introduce regularization, in this example having a similar effect as the L$_2$ term in Eq.~(\ref{eq:ridge}).  

The Bayesian viewpoint has found wide use in ML. Later we will see a second example of Bayesian methods applied to  global optimization.

\subsection{Artificial neural networks}

Having taken a pass through the general mechanics of ML regression, we now turn to a more complex model type: artificial neural networks (ANNs).  ANNs are among the most commonly used ML models, now almost synonymous with ML to the public.  This course does not have the scope for a deep dive into ANNs, but it is instructive to apply the formalism from Section~\ref{sec:lin_reg} to a new type of model.  

ANNs were inspired by biological nervous systems.  The base component is the neuron, which consists of three components: input signals ($\vx$), weights on each input (usually written $\vw$ but playing the~same role as $\vtheta$ in linear regression), and an activation function $f$, which combines the inputs and weights to produce an output $a$.  Note that typically the bias term $b$ is specified explicitly rather than the~implicit $\theta_0$ in linear regression. We can then write the neuron's output as $a = f(\vx,\vw,b)$. The activation $f$ can be as simple as a linear function: in this case, the task of fitting a single neuron looks just like the~regression task from the first section. Typically, non-linear functions such as a sigmoid or Tanh are used to model more complex behavior.  A common choice of activation function is the rectified linear unit (ReLU), which outputs a linear function for positive inputs and zero for negative inputs.  

\begin{figure}[!htb]
\centering
\includegraphics[width=.6\linewidth]{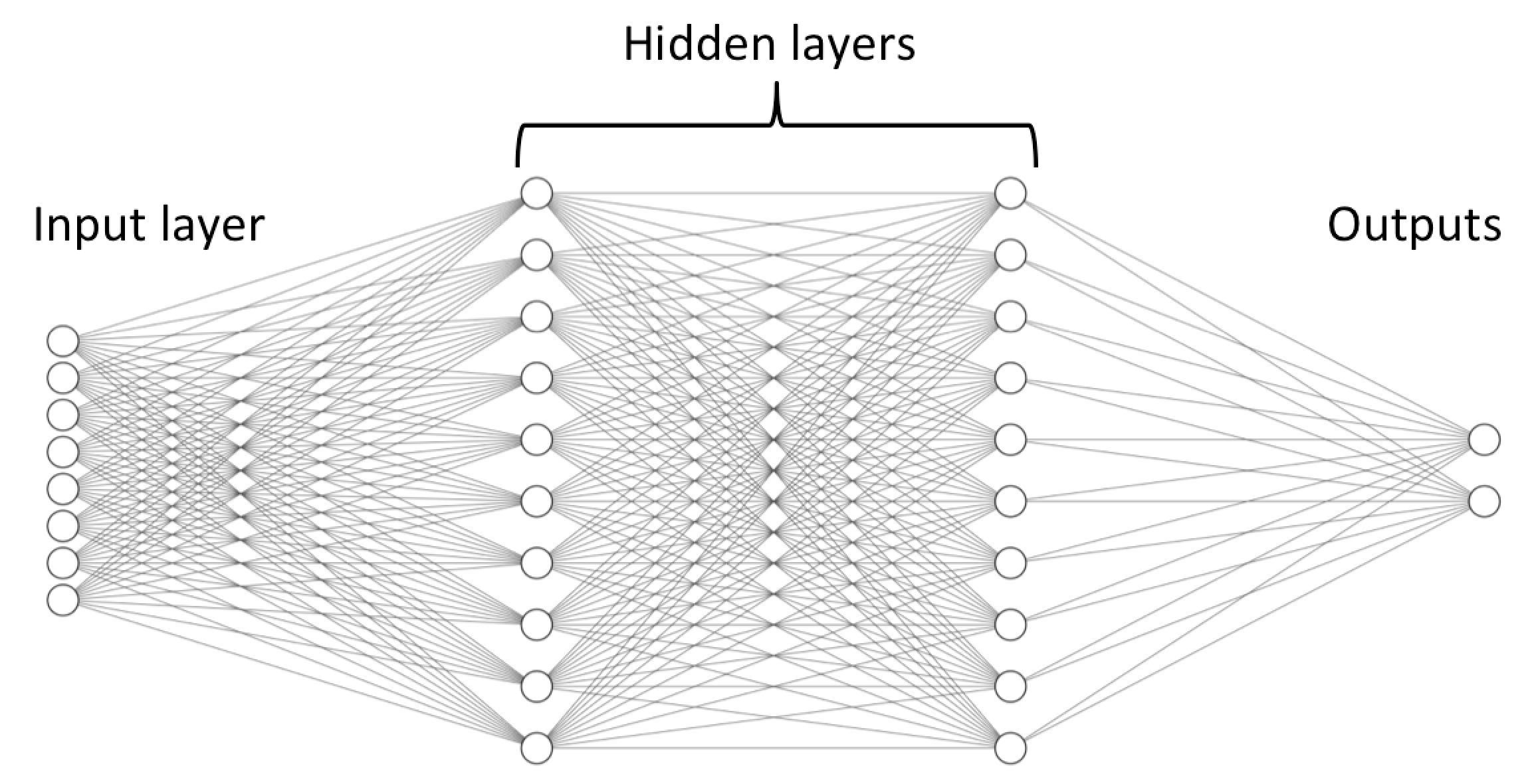}
\caption{Schematic of a fully-connected ANN with eight input features, two output labels, and two hidden layers with ten neurons each.  (Figure courtesy Alex LeNail). }\label{fig:ANN}
\end{figure}

Linking together multiple neurons, e.g., such that one layer's output is the next layer's input (Fig.~\ref{fig:ANN}), creates an ANN.  The first layer's inputs are the training set features and the final layer outputs the prediction, while any intermediate layer is called a `hidden' layer.  As in linear regression, training the~network requires a cost/loss function, $C_{\vw,b}(\vx)$, that calculates the difference between the~output layer and the training labels for any choice of $\vw,b$.  There is no closed-form solution analogous to the normal equations, so training uses gradient descent,
\begin{align}
w_j := w_j - \alpha \frac{\partial C_{\vw,b}}{\partial w_j} \,,
\label{eq:nnw}
\end{align}
with
\begin{align}
\frac{\partial C_{\vw,b}}{\partial w_j}  \approx \frac{C_{\vw + \vect{\epsilon},b} - C_{\vw,b}}{|\vect{\epsilon}|}  \,.
\label{eq:nndc}
\end{align}
There is one complication here worth noting: with $n$ weights, each update requires $n$ calculations of Eq.~(\ref{eq:nndc}), and each calculation requires a full forward pass through the network (also $\bigO(n)$), so each model update is $\bigO(n^2)$.  With $n$ of order millions for large networks, training would be prohibitively computationally expensive.  Luckily, there is another approach, using the chain rule to calculate the individual gradients for each parameter.  Because the method starts at the output layer and moves back towards the input layer it is known as `backpropagation.'  While at first glance, this would appear even less efficient than Eq.~(\ref{eq:nnw}), it is possible to express the gradients such that the chain rule components are shared.  Consequently the backpropagation update requires only a single pass forward and then backwards through the network, with $\bigO(n)$ computations.  For a derivation, the reader is referred to e.g. \cite{nielsenNN}.

Training an ANN involves many of the same considerations as the simple linear model.  Expanding the number of features or adding nodes and connections increases the power of the model, but also increases the risk of overfitting.  As in linear regression, if loss on the training set significantly outperforms the validation set, imposing L$_1$ or L$_2$ norms on the fitting parameters reduce model variance. For ANNs, there are additional regularization techniques, such as adding noise at the input layer or randomly blocking a selection of neurons (known as 'dropout') during training.

The choice of ANN architecture, i.e. the pattern of connections between neurons, depends on the~problem type. In simple fully-connected networks, e.g. Fig.~\ref{fig:ANN}, all nodes in adjacent layers share connections. However, when there are a large number of features, e.g. for images, fully connected networks may require an unmanageable number of parameters. Instead, convolutional neural network (CNNs) use only a small number of local connections, which are then convolved over a larger image. CNNs naturally look for local features in the image (e.g. edges) that can be combined to form abstract concepts in later layers. Similarly, for sequential processes, e.g. natural language processing or time-series data, recurrent architectures (RNNs) naturally capture temporal patterns.  The term `deep learning' describes network architectures with many hidden layers: the early layers effectively play the role of feature engineering, while later roles process the data into more complex quantities for further abstraction. In recent years, deep learning with CNNs and RNNs has become a field unto itself.

\subsection{Logistic Regression}

\begin{figure}[!htb]
\centering
\includegraphics[width=.6\linewidth]{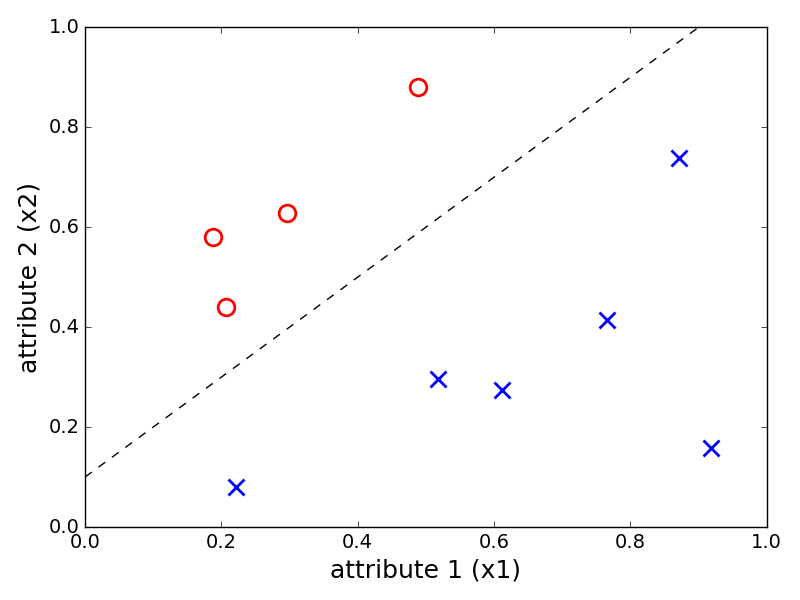}
\caption{Two-class classification problem, for example predicting if a set of parameters is safe (blue crosses) or will cause a beam trip (red circles).  The goal is to find a decision boundary, e.g. dashed black line, that separates trips from safe operation.}\label{fig:class}
\end{figure}

We now turn to a new type of problems common to ML: classification.  Rather than predicting a continuous variable as in regression, we instead predict class membership.  For example, consider predicting whether a set of parameters will cause a machine trip (Fig.~\ref{fig:class}). We could still use a regression model, with labels $y= [0,1]$, and interpret the prediction, $h_\theta(\vx)$ as a probability of a trip. But how are we to interpret predictions of $h_\theta(\vx) < 0$ or $h_\theta(\vx) > 1$?  Instead, consider the addition of the logistic function, $g(z) = 1/(1+e^{-z})$ to give
\begin{align}
h_\theta(\vxi) = g(\vxi\cdot\vtheta) = \frac{1}{1+\exp(-\vxi\cdot\vtheta)} \,.
\end{align}
The hypothesis is now constrained to be on the interval $(0,1)$.  Due to the inclusion of the non-linear $g(z)$, the normal equations (Eq.~(\ref{eq:normal})) are no longer applicable, but applying MLE still gives an update rule
\begin{align}
\theta_j := \theta_j + \alpha \sum_{i=1}^m \big[\yi - h_\theta\big(\vxi\big)\big] \xi_j \,.
\label{eq:log_update}
\end{align}
The logistic update is identical to the linear regression update, except that $h_\theta(\vxi)$ is now non-linear.  Indeed it is possible to treat both problems as sub-classes of the generalized non-linear model (see Chapter~1, Section III from \cite{cs229}).

Evaluating the quality of a classification model requires some care. While the cost function gives a~relative score during training, it is not easily interpretable. One tempting choice is accuracy, i.e. the~fraction of correctly classified examples.  However, accuracy tells us nothing about the distribution of false positives vs. false negatives.  For an extreme case consider a highly uneven class distribution, with 99\% negative and 1\% positive examples. A trivial model $h_\theta(\vx) = 0$ has the impressive seeming accuracy of 99\%, and yet has zero predictive power based on the input features. A better metric is the~paired combination of precision/recall, with `recall' the~fraction of true events identified, and `precision' the~fraction of predicted true events that are correct.  Our trivial model of $h_\theta(\vx) = 0$ has a recall of zero (0\% of events found) and an undefined precision (zero out of zero events correct), and thus is clearly not an effective model. 

\begin{figure}[!htb]
\centering
\subfigure[]{
\includegraphics[width=.45\linewidth]{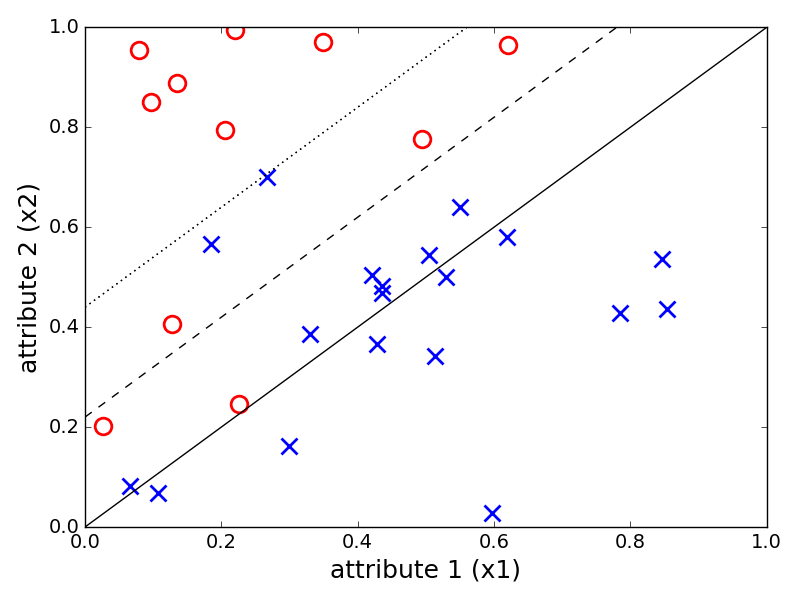}}
\subfigure[]{
\includegraphics[width=.45\linewidth]{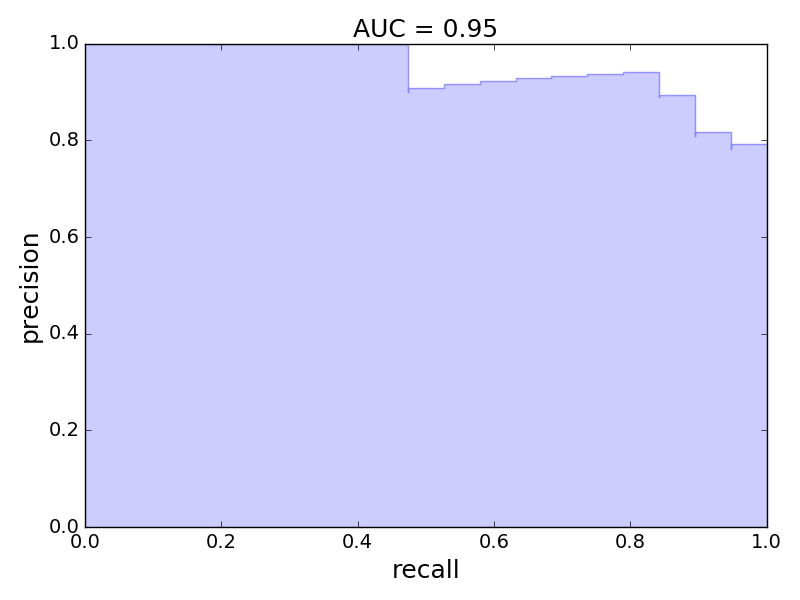}}
\caption{We now consider a noisier classification case (left), with the trips (again red circles) no longer separable by a linear model. Depending on the application, the user may want a strict model (solid line) that identifies all trips but has many false positives, or a weaker alarm (dot-dashed line) that avoids unnecessary warnings but misses some trips. The precision-recall curve (right) captures this trade-off for a logistic regression model trained on the~same data.  The area under the curve (AUC), i.e., the blue region, condenses the performance into a single scalar score.}\label{fig:class_noise}
\end{figure}

Secondly, logistic regression gives a probability score, rather than a boundary; the user must select a threshold to draw the boundary itself. Consider the case of Fig.~\ref{fig:class_noise}(a) with three possible boundaries.  For the given data, no linear model perfectly separates the two classes. The boundary preference depends on the application: for example in a machine protection system the user may wish to weigh the danger of missing a true positive (leaning towards high recall) with the annoyance of constant trips from false warnings (leaning towards high precision) depending on the severity of the trip.  Consequently, the user may want to know the precision and recall for a range of thresholds.  To condense the score to a~single number, it is common to plot the tradeoff between precision and recall, and report the area under the~curve (AUC) (Fig.~\ref{fig:class_noise}(b)).  It is then up to the user to select the preferred threshold.  AUC is also commonly applied to the receiver operator characteristic (ROC), an alternative metric pair used when class probabilities are roughly even.

\section{Non-parametric models}

\begin{figure}[!htb]
\centering
\includegraphics[width=.6\linewidth]{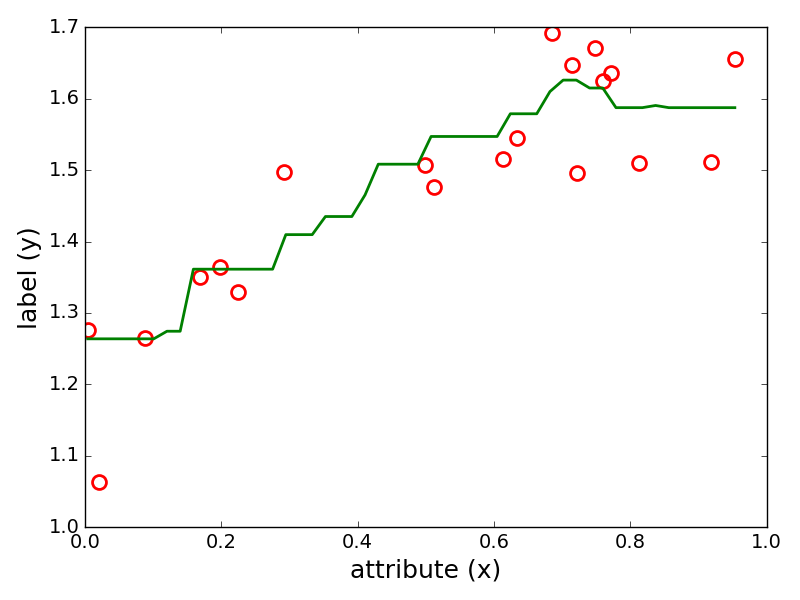}
\caption{We revisit the problem of Fig.~\ref{fig:linreg}. This time we fit the data (red circles) with a KNN model with $k=5$ (green line).}\label{fig:knn}
\end{figure}

To this point, we have only considered parametric models of the form $h_{\theta_1,...,\theta_n}(\vx)$, with explicitly defined fitting parameters.  We now turn to non-parametric models, $h_{x^{(1)},...,x^{(m)}}(\vx)$, where the model itself is built on instances in the training set.  (For this reason non-parametric models are also described as `instance-based learning.')  As a simple illustration, consider a model in which a prediction is given by the value of the nearest example
\begin{align}
y^{(j)} = y^{(i^*)}\,, \,\, i^* = \underset{i}{\mathrm{argmin}} ||\vx^{(j)}-\vx^{(i)}|| \,.
\label{eq:knn}
\end{align}
Equation~(\ref{eq:knn}) is a subset of the popular k-nearest neighbors (KNN) model, with $k=1$; in general, the prediction is given by an average over the $k$ nearest neighbors. Figure~\ref{fig:knn} shows a KNN applied to the~regression problem from the beginning of the write-up. Though simple, KNNs can be very effective and  are popular in industry. 

\begin{figure}[!htb]
\centering
\subfigure[]{
\includegraphics[width=.48\linewidth]{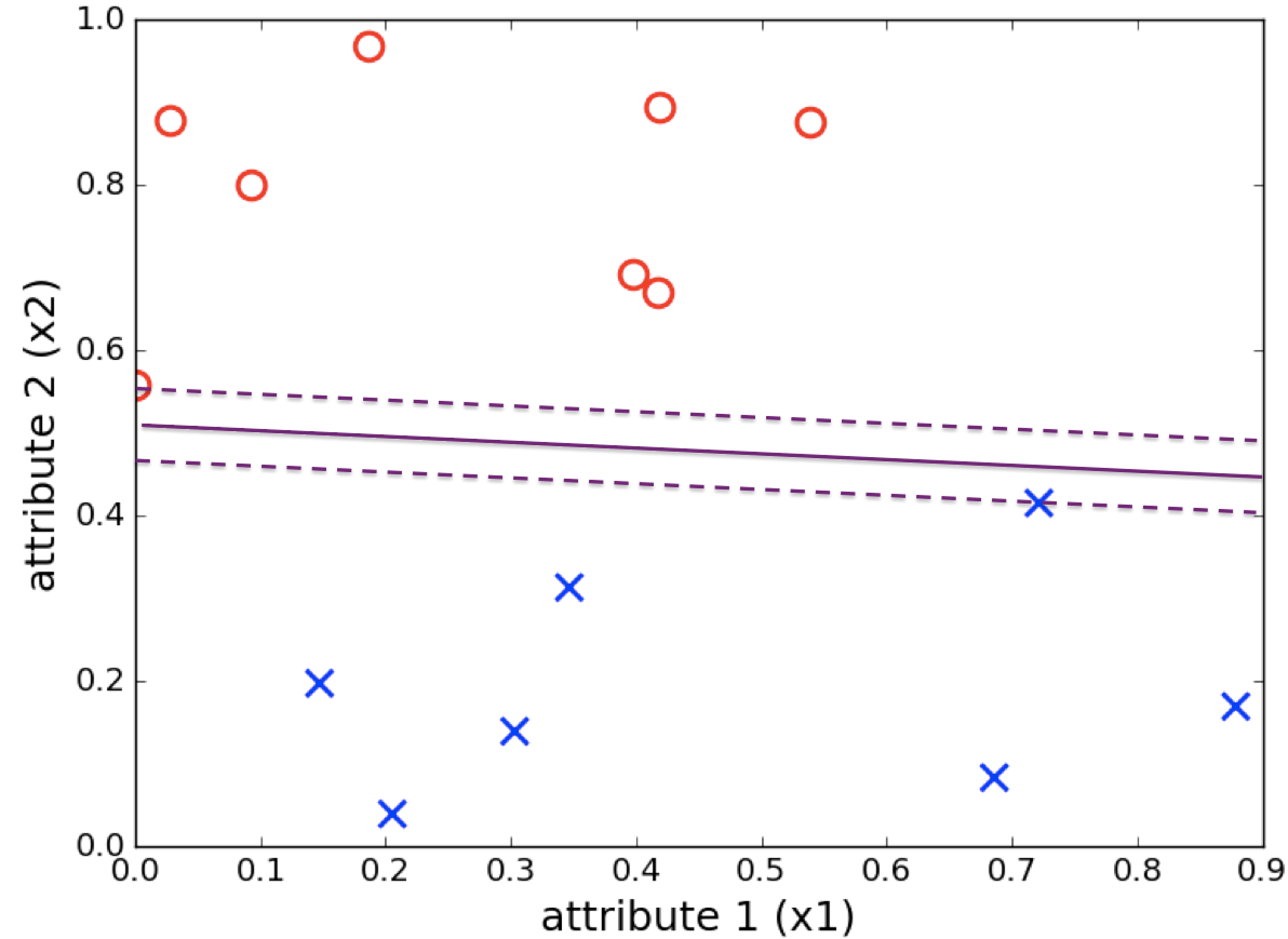}}
\subfigure[]{
\includegraphics[width=.48\linewidth]{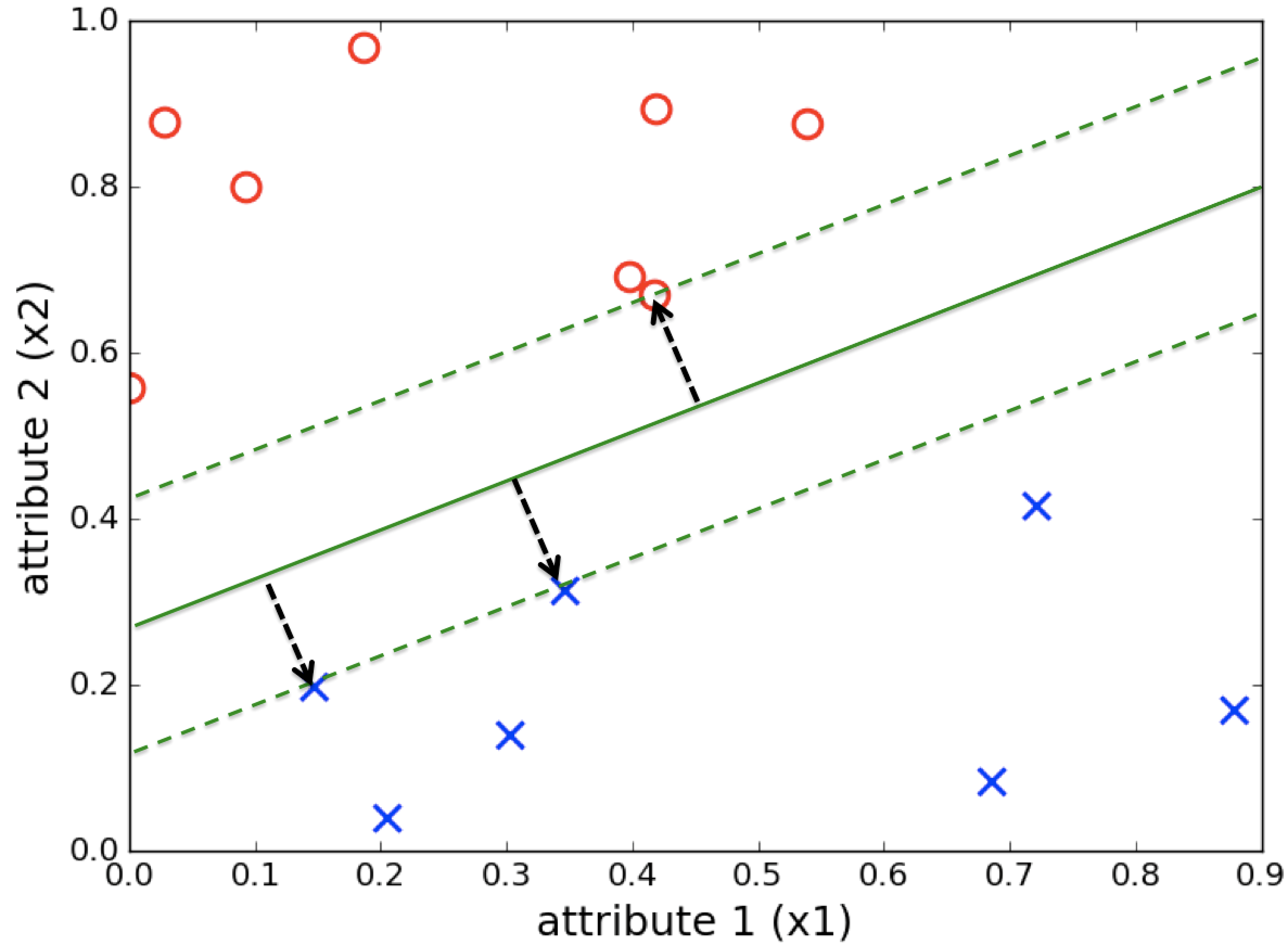}}
\caption{In the two plots above, two different decision boundaries (solid lines) both perfectly divide the examples into two classes (red circles, blue crosses). An optimal margin classifier prefers the boundary on the right because the closest example to the boundary (the max-min distance) is larger.  Equivalently the boundary at right has a~larger margin, i.e., separation of the dashed lines. For an SVM, the `support vectors' (black arrows on right) define the~boundary.}\label{fig:omc}
\end{figure}

Non-parametric models are also applied to classification problems. One example is the optimal-margin classifier. For a pictorial understanding, consider the case of Fig.~\ref{fig:omc}: two possible decision boundaries both perfectly classify the training examples, but we may intuitively prefer the solid line.  The~margin classifier quantifies this intuition by selecting the line that maximizes the distance from the decision boundary to the nearest instance. 

The support vector machine (SVM) is the most famous example of a margin classifier.  The SVM chooses a boundary surface, defined by parameters $\vw, b$ by solving the minimization problem
\begin{align}
\underset{\vw,b}{\mathrm{min}} \,\,\, ||\vw||^2 \,\,\,
\mathrm{s.t.} \,\,\, \yi(\vxi\vw + b) \geq 1 
\end{align}
for all examples $i$ in the training set. (Here we again use the ANN notation, also popular for SVMs, with $\vw$ in place of $\theta$ for $j>0$, and explicit bias term $b=\theta_0$.) The derivation of both this optimization constraint and the resulting solution is beyond the scope of these notes, but it is an interesting application of duality in optimization and worth a close read for the dedicated student  (see e.g. Ref. \cite{cs229} Chapter 3).  Here we simply state the result: having solved for the optimal parameters, $\alpha_i$, of the dual problem, we make a prediction for a new point $\vx'$ from
\begin{align}
h(\vx') = \mathrm{sgn}(\vx'\vw+b) 
= \mathrm{sgn}\bigg( \sum_{i=1}^m \alpha_i \yi \langle \vxi, \vx' \rangle + b \bigg) \,.
\label{eq:svm}
\end{align}
Most of the $\alpha_i$ will tend to zero, and only a small number of examples (the eponymous `support vectors') are needed to calculate Eq.~(\ref{eq:svm}) during inference, making the models computationally tractable. The~reason we write out Eq.~(\ref{eq:svm}) is to highlight one critical point: in both the definition of the dual problem (not shown) and the inference procedure for new points (Eq.~(\ref{eq:svm})), the examples $\vxi, \vx'$ only enter the calculation through an inner product $\langle \cdot, \cdot \rangle$.  In the next section we will see the importance of this observation.  

\pagebreak
\subsection{Kernel Trick}

\begin{figure}[!htb]
\centering
\subfigure[]{
\includegraphics[width=.45\linewidth]{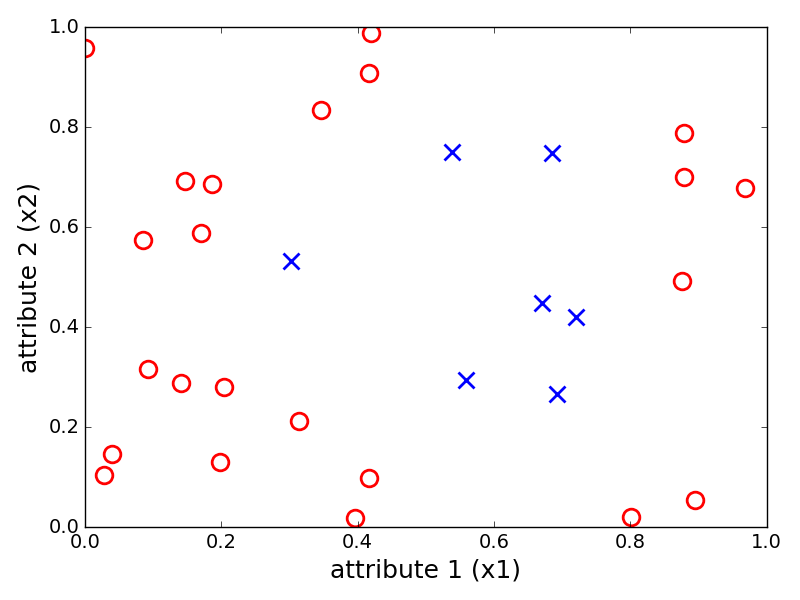}}
\subfigure[]{
\includegraphics[width=.52\linewidth]{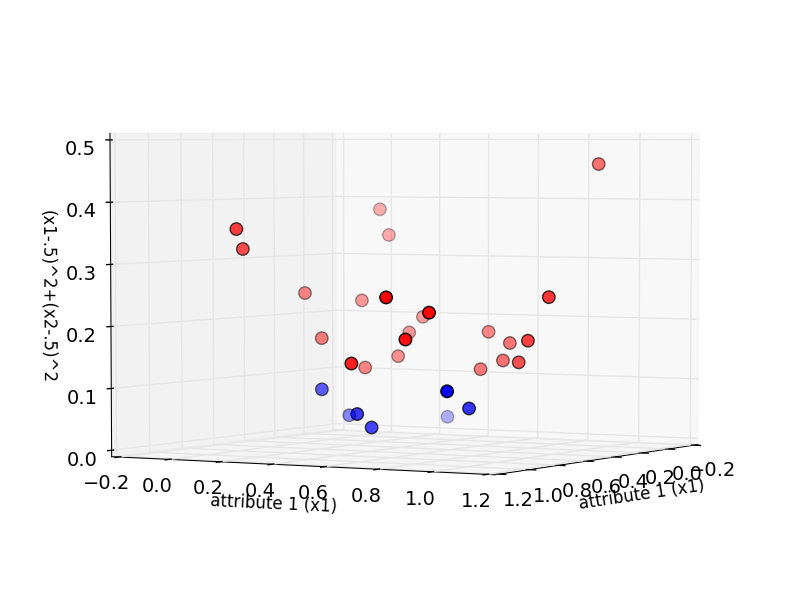}}
\caption{Left, a training set that is not separable by a linear function (left).  Adding new features, $x_1^2 + x_2^2$ lifts the~problem into a higher dimensional space where the examples are now separable by a linear surface.}\label{fig:class_circ}
\end{figure}

The examples of Fig.~\ref{fig:omc} were separable by a linear boundary, but now consider the case of Fig.~\ref{fig:class_circ}.  As with linear regression, we can use feature generation to introduce non-linearities to the model; in the~case of Fig.~\ref{fig:class_circ}, adding a new feature of the form $x_1^2 + x_2^2$ `lifts' the problem into a higher dimensional space in which the problem is linearly separable.  Or using the notation of Eq.~(\ref{eq:feat_def}) we have introduced the mapping $x_1, x_2 \rightarrow \vphi(x_1,x_2) = \{x_1,x_2, x_1^2 + x_2^2\}$.

Unfortunately, feature generation can also be computationally expensive.  Here we introduce a~subtle but powerful alternative known as the kernel method. Let's return to our observation that Eq.~(\ref{eq:svm}) is expressed entirely in terms of inner products.  Incorporating feature mapping into Eq.~(\ref{eq:svm}) gives us a new SVM inference equation
\begin{align}
\vphi(\vx')\cdot\vw+b = \sum_{i=1}^m \alpha_i \yi \langle \vphi(\vxi), \vphi(\vx') \rangle + b \,.
\label{eq:svm_fm}
\end{align}
(Of course we also have to resolve the dual problem to find the new $\alpha_i$.) Next we define a kernel function, $K(\vxi,\vx') \equiv \langle \vphi(\vxi), \vphi(\vx') \rangle$.  For discrete observations, we write this function as $\vxi K (\vx')^T$, for some square matrix $K$. We now have an alternative formulation for the inference equation
\begin{align}
\vphi(\vx')\cdot\vw+b = \sum_{i=1}^m \alpha_i \yi  K(\vxi,\vx') + b \,.
\label{eq:svmK}
\end{align}
Comparing Eqs.~(\ref{eq:svm_fm}) and (\ref{eq:svmK}) it may appear we have added a trivial piece of formalism.  However, a~result due to Mercer makes this subtle change deceptively powerful. Rather than explicitly choosing a mapping, $\vphi(\vx)$, and then calculating the corresponding $K$, Mercer's theorem tells us we are free to choose any positive semi-definite $K$, and we can skip the step of explicitly calculating $\vphi(\vx)$. To appreciate the~advantage of the kernel method, consider the kernel $K(\vx^{(1)},\vx^{(2)}) = (\vx^{(1)} \cdot \vx^{(2)})^2$, which corresponds to the mapping $\phi(\vx) = \{x_i x_j \, | \, i,j \in n \}$ for an $n$-dimensional vector $\vx$ (see SVM chapter in Ref. \cite{cs229}).  Though the corresponding models (Eq.~(\ref{eq:svm_fm}) and (\ref{eq:svmK})) are identical, the Kernel method has complexity $\bigO(n)$ while the direct mapping has complexity $\bigO(n^2)$.  The gain can be dramatic: the popular squared exponential (SE) kernel $K(\vx^{(1)},\vx^{(2)}) = e^{-||\vx^{(1)}-\vx^{(2)}||^2}$ corresponds to an infinite dimensional mapping. Simply put, the kernel method provides the model complexity of a high-dimensional mapping without the computational overhead.

The kernel method is applicable to any instance-based model in which the training data enter only as inner products.  A second common example is the Gaussian process (GP). One appeal of GPs is the~convenient encoding of uncertainty prediction, which makes them particularly useful for describing scientific problems.  For the same reason, GPs are commonly used in Bayesian optimization; see Refs.~\cite{mcintire16ce,kirschner19mh} for application to accelerator optimization. Rasmussen (Ref. \cite{rasmussen06w}) is recommended for a thorough introduction to GPs.

\section{Other types of machine learning}

Introductions to machine learning commonly divide the field into three distinct branches: supervised learning, unsupervised learning, and reinforcement learning.  While this note focuses primarily on supervised learning (the most widely used of the three), in this section we briefly cover the other two branches.

\subsection{Unsupervised learning}

In supervised learning the training set consists of both input features, $\vx$, and labels, $y$ (hence the term `supervised'). We now consider unsupervised learning, in which case the training set consists only of the~input features, $\vx$.  For example, consider the challenge of dividing a training set into similar groups based on shared characteristics. When the examples are labelled, this is a supervised classification problem.  However, even without labels, we can still group the examples by self-similarity. Indeed it is not even necessary to know the number of classes. Examples of clustering algorithms include K-means (note no-relation to KNN), density-based spatial clustering of applications with noise (DBSCAN), Gaussian mixture models (GMMs), and hierarchical clustering.  

A second common type of unsupervised learning is anomaly detection, i.e., identifying outliers in a set of examples.  A related challenge is breakout/changepoint detection, which looks for changes in sequential data.  For example, consider a time series of a vacuum pump; a single spurious high value (e.g.~a~faulty reading) would be an anomaly, whereas a shift to a new average level (e.g. due to a leak) would be a breakout.  Both problems can make use of clustering algorithms, as well as modified ANNs and SVMs among other algorithms. Yet another task is decomposition of a signal into its components. A classic example is the 'cocktail problem' of separating voices in a recording of a cocktail party. Independent component analysis (ICA) is a popular decomposition algorithm.

\subsection{Reinforcement learning}

Reinforcement learning (RL) is inspired by human learning.  In both supervised and unsupervised learning, the training data is defined prior to training.  By contrast, in reinforcement learning (RL) the training set is generated dynamically by interaction with an environment during the learning process.  In RL, an~agent exists in an environment, consisting of multiple states, which are connected by actions.  Given a state, $s$, the agent chooses an action, $a$, resulting in a new state $s'$ (either deterministically or stochastically).  Finally, the agent receives rewards or penalties based on its path through the environment.  The~goal of the agent then is to find the optimal `policy,' i.e. the action associated with each state that maximizes the long-term rewards.  As an example, consider playing the game checkers: given a particular position in the game, $s$, the agent moves one of the pieces, which is the action $a$.  Following interaction with the environment, i.e., the opponent moves, the agent is presented with a new game position, $s'$.  The~agent may get periodic awards (e.g. capturing a piece), or may be given a single award at the end of the game for winning or losing. Finally, the agent updates the policy, $\pi(a,s)$, based on the rewards. So called `deep' RL is an increasingly popular variant using ANNs to reduce the dimensionality of the state and/or action space. An example is AlphaZero, as of late 2018 arguably the best Go player in the world.  For interested readers, Ref. \cite{sutton98b} by Sutton and Barto is recommended for a thorough introduction to RL.

\section{Examples from accelerator physics}

Machine learning applications are increasingly popular throughout physics (see e.g. Ref. \cite{radovic18wr} for a recent review for particle physics).  Accelerator physics is no different, with a long history of applications and a~growing enthusiasm in the last few years (see e.g. Ref. \cite{ICFAML18} for a recent summary).  

The CAS presentation walked through a few specific applications of ML to x-ray free-electron lasers (XFELs).  The first application presented was analysis of two-dimensional diagnostics.  For example images of the longitudinal phase space from an x-band transverse deflecting cavity (XTCAV) are critical for both optimizing FEL performance and also as a user diagnostic \cite{behrens14etal}. Preliminary results show CNNs can outperform state-of-the-art hand-written algorithms in complex analysis of the XTCAV. ANNs are also useful for solving inverse problems; rather than rerunning iterative solvers from scratch for each new example, ANNs are trained to learn a general solution in an offline training process and then run online inference on each new example in a fraction of a second. An application to astrophysics found a speed-up factor of 10 billion \cite{hezaveh17lm}.

Bayesian optimization applies the concept of Bayes' rule to global optimization problems. As opposed to model-independent strategies, for example gradient descent, Bayesian optimizers construct a~model of the target system.  The model conveys two benefits: first, an `acquisition function' weighs the~predictions and uncertainties of the model to suggest the most valuable next point to measure (balancing the `exploration-exploitation tradeoff'). Second, the model can be trained on previous data, simulations, and theory, providing additional guidance for the search. While Bayesian methods have high computational complexity, in accelerator applications the computation time is often negligible compared to the~sampling time.  As noted earlier, Bayesian GP optimizers have been used successfully for online tuning of XFELs \cite{mcintire16ce,kirschner19mh}.

The final example showed how regularization speeds convergence of ghost imaging (GI); formulating GI as a linear regression problem \cite{li18ck, ratner19cl} enables use of compressed sensing \cite{candes06rt}.  Other examples of machine learning in FEL physics briefly mentioned included tuning with reinforcement learning \cite{wu}, building fast surrogate models to mimic high-fidelity simulations \cite{edelen}, and diagnosing beam trips with multi-variable anomaly detection. For more examples, a summary of the first ICFA workshop on ML gives a broad overview of applications to accelerators \cite{ICFAML18}.

\section{Tips for training models}

Finally, we conclude with brief practical advice to the first-time machine-learner. The first, and often the most difficult, step of ML is assembling the training data set.  It should be expected that collecting or generating high quality data will take more time than training itself. While training sets commonly consist of collected/measured data, the prevalence of high-fidelity models in accelerator physics may make training from simulations feasible as well.  The number of training examples required depends greatly on the problem complexity; checking performance vs. fraction of the data set used for training can help determine if more examples are needed.

Machine learning methods are highly effective at interpolation, but typically fail at extrapolation; whether measuring or simulating data, care should be taken that the training set encompasses the parameter range encountered during inference on real examples. Cleaning the data set to remove outliers or anomalous conditions is also critical. Dimensionality reduction, i.e. removing redundant or irrelevant features, can reduce both training time and overfitting.

Scientific problems pose unique challenges for ML.  The need for model interpretability and robustness may push a science applications towards simpler model types or architectures. Secondly, standard assumptions of feature independence and noise may not hold in physics problems; students are advised to check the underlying assumptions of the chosen model before applying to accelerator data. For example when collecting data sets from measurements, accelerator diagnostics may introduce significant noise on both the dependent and independent variables.  The latter violates standard assumptions for even least squares regression, leading to regression dilution.  (For an example, see ghost imaging \cite{ratner19cl}.) Scientific data types, e.g. predicting complex-valued functions, can also require customized solutions. 

Finally, we end with a personal opinion of this author: while `deep learning' from raw data has an~understandable allure, as of early 2019 careful consideration of the physics, both in feature engineering and selection of model architecture, is still worth the extra attention. Happy learning!

\bibliographystyle{unsrt}

\end{document}